\documentclass[english]{article}
\usepackage[T1]{fontenc}
\usepackage[latin1]{inputenc}
\usepackage{color}
\usepackage{amssymb}

\makeatletter

\providecommand{\tabularnewline}{\\}

\usepackage{color}

\usepackage{setspace}



\AtBeginDocument{
  
}

\makeatother

\usepackage{babel}

\begin{document}

\title{\textbf{\normalsize Study of Chemical Equilibration of a Baryon Rich
QGP}\\
\textbf{\normalsize }\\
\textbf{\normalsize }\\
\textbf{\normalsize }\\
}

\author{{\normalsize Abhijit Sen}\\
{\normalsize Lecturer, Department of Physics, Suri Vidyasagar College,
Suri-731101, INDIA}\\
{\normalsize }\\
{\normalsize{} }\\
{\normalsize }\\
{\normalsize }\\
{\normalsize{} }\\
{\normalsize }\\
{\normalsize }
{\normalsize }\\
{\normalsize }\\
{\normalsize }\\
{\normalsize }\\
{\normalsize PACS No. 12.38.Mh(Quark Gluon Plasma), 21.65(Quark Matter),25.75(Relativistic
heavy ion collisions)}\\
{\normalsize }\\
{\normalsize }\\
{\normalsize }\\
{\normalsize }\\
}

\maketitle
\begin{abstract}
{\normalsize Parton equilibration studies for a thermally equilibrated
but chemically non-equilibrated quark-gluon plasma (QGP) , likely
to be formed at the relativistic colliders at BNL and CERN is presented.
Parton equilibration is studied enforcing baryon number conservation.
Process like quark - flavour interchanging is also taken into consideration.
The degree of equilibration is studied comparatively for the various
reactions / constraints that are being considered. }{\normalsize \par}
\end{abstract}

\section{Introduction}

\textcolor{black}{The parton plasma likely to be created in high energy
heavy ion collisions is sure to be in a state of both thermal and
chemical non-equilibria.Once thermal equilibrium is attained, the
subsequent evolution takes place according to the laws of hydrodynamics.}

\textcolor{black}{It is known that {[} 1 ] the initial system produced
at RHIC energies have a finite non- zero baryon density. The anti-particle
to particle ratios at mid-rapidity for $\sqrt{s_{NN}}$=130 GeV Au+Au
collisions at the STAR collaboration,BNL report that there is a noticiable
excess of baryons as compared to anti-baryons as reflected by the
yields of $\frac{{\displaystyle \overline{p}}}{{\displaystyle p}}$
( hovering between 0.6 and 0.8 ),$\begin{array}{c}
\frac{{\displaystyle \overline{\Lambda}}}{{\displaystyle \Lambda}}\end{array}$( hovering between 0.7 and 0.8 ) {[} 2 ] for varying transverse momentum,
centrality and rapidity. In a subsequent reporting of the BRAHMS collaboration
of BNL{[} 3 ] it is seen that during Pb+Pb collisions at $\sqrt{s_{NN}}$=
200 GeV the observed $\begin{array}{c}
\frac{{\displaystyle \overline{p}}}{{\displaystyle p}}\end{array}$ ratio is a maximum of around 0.8 at zero rapidity and falls steadily
for higher rapidity values. }

\textcolor{black}{This excess of baryons ( and hence quarks ) over
anti-baryons ( and hence anti-quarks ) clearly necessiates the inclusion
of a chemical potential into the theoretical framework . Moreover,
strong interaction conserves baryon number, which also must be incorporated
in our scheme. The present piece of work is aimed at attaining such
an objective. }\\
\textcolor{black}{}\\
\textbf{\textcolor{black}{\large }}\\
{\large \par}

\section{\textcolor{black}{\large Brief Review of Previous Works}\protect \\
}

.

\textcolor{black}{Inclusion of chemical potential into the framework
of Parton equilibration studies has been rather recent. Although in
1986 Matsui et.al reported {[} 8 ] on strangeness production rates
at nonzero chemical potential, it was not until the very end of the
past century that a group from BARC,Mumbai, India included a chemical
potential into chemical equilibration studies {[} 4,5 ]. They considered
a chemical potential (for massless quarks) that equalled the system
temperature at all temperatures which was not realistic as it did
not consider baryon number conservation. Furthermore, they did not
distinguish between quarks and anti-quarks. The same criticism regarding
chemical potential and baryon number conservation also applies to
an attempt {[} 6 ] (incomplete) by the present author during the preliminary
stages of the current work. However, explicit masses and distinction
between quarks and anti-quarks had been incorporated in that conference
report. In 2004, He et.al. {[} 7 ] undertook a much more complete
study including baryon number conservation. They used an expansion
of the number densities over powers of chemical potential. We basically
follow the same line with a few points in difference with the work
by He et.al.The points of difference are:}

\textcolor{black}{i) We explicitly introduce the strange quark mass}

\textcolor{black}{ii) We treat quarks and antiquarks separately.It
can be noted here that although He et. al. {[} 7 ] started off with
quark-antiquark distinction, subsequently that distinction was put
away. }

\textcolor{black}{iii) We include results of full phase space calculations
for both quark-antiquark pair production and quark flavour changing
processes.}{\large }\\
\textbf{\large }\\
{\large \par}

\section{\textcolor{black}{\large The Chemical Potential}}

\textcolor{black}{At this point let us note the following points:}\\

\textcolor{black}{1. The initial strangeness content of the QGP fireball
is zero and moreover strangeness is very nearly conserved in strong
interaction which clearly indicates that the strange quark chemical
potential must be zero.}

\textcolor{black}{2. If we consider a pair production process in QGP
( production of non-strange quark- antiquark pair) it can be argued
that the net chemical potential on both sides of the reaction must
be the same. Since gluons have no chemical potential clearly this
would imply }

\[
\mu_{q}=-\mu_{\overline{q}}\]
$\qquad\qquad\qquad\qquad\qquad\qquad\qquad\qquad\qquad\qquad\qquad$...(1)

\textcolor{black}{3. We treat the up and down flavours in the same
footing and treat them as massless. }

\textcolor{black}{4. For a given initial baryon number density and
initial values of the light quark \& antiquark non-equilibrium fugacities
( which is defined as the ratio of the number density to the equilibrium
number density) we can find the initial light quark chemical potential
by method of iteration.}

\textcolor{black}{5. The rate of change of chemical potential with
time is dictated by the baryon number conservation equation.}\\

\section{\textcolor{black}{\large Thermodynamic \& Statistical Mechanic Treatments}\textcolor{black}{\normalsize }\protect \\
}

\subsection{\textcolor{black}{\normalsize Distribution functions}\protect \\
}

\textcolor{black}{For a thermally equilibrated but chemically non-equilibrated
parton plasma, the distribution functions are given by the Juttner
distribution functions {[} 9 ]. In the present paper, initially we
use the full distributions. In the latter stage we shall adopt certain
approximations in order to achieve some mathematical simplicity as
also to confront some mathematical unsolvability that we shall encounter.
The full Juttner distribution functions are }\\

\textcolor{black}{$\begin{array}{c}
{\displaystyle f_{g}=\frac{\lambda_{g}}{e^{\varepsilon/T}-\lambda_{g}}}\\
{\displaystyle f_{q(\overline{q})}=\frac{\lambda_{q(\overline{q})}}{e^{(\varepsilon\mp\mu_{q})/T}+\lambda_{q(\overline{q})}}}\\
{\displaystyle f_{s}=f_{\overline{s}}=\frac{\lambda_{s}}{e^{\varepsilon/T}+\lambda_{s}}}\end{array}$}$\qquad\qquad\qquad\qquad\qquad\qquad\qquad\qquad$...(2)\\

\subsection{\textcolor{black}{\normalsize Basic Quantities}\protect \\
\protect \\
}

\textcolor{black}{Using standard techniques, we can obtain the various
thermodynamic quantities that are required to specify the state of
the system. We obtain:}\\
\textcolor{black}{$\begin{array}{c}
{\displaystyle Q=Q_{g}+}{\textstyle \Sigma_{i=u,d,s}(Q_{i}+Q_{\overline{i}})}\end{array}=Q_{g}+2*(Q_{q}+Q_{\overline{q}})+2Q_{s}$} $\qquad\qquad$...(3)\textcolor{black}{}\\
\textcolor{black}{ Q being the number density, energy density or pressure
in general and q denoting the massless quark flavours. }\\
\textcolor{black}{This gives }\\
\textcolor{black}{$\begin{array}{c}
{\displaystyle n=\{\frac{16T^{3}}{\pi^{2}}\Sigma_{k=1}^{\infty}\frac{\lambda_{g}^{k}}{k^{3}}+\frac{12T^{3}}{\pi^{2}}\Sigma_{k=1}^{\infty}(-1)^{k-1}\{\frac{\lambda_{q}^{k}e^{\frac{k\mu_{q}}{T}}+\lambda_{\overline{q}}^{k}e^{-\frac{k\mu_{q}}{T}}}{k^{3}}\}+}\end{array}$}\\
\textcolor{black}{$\qquad\qquad\qquad$${\displaystyle \frac{6m_{0s}^{3}}{\pi^{2}}\Sigma_{k=1}^{\infty}(-1)^{k-1}\lambda_{s}^{k}\frac{K_{2}(kx_{s})}{(kx_{s})}\}}$}\\
\textcolor{black}{$\begin{array}{c}
{\displaystyle \begin{array}{c}
{\displaystyle \varepsilon=\{\frac{48T^{4}}{\pi^{2}}\Sigma_{k=1}^{\infty}\frac{\lambda_{g}^{k}}{k^{4}}+\frac{36T^{4}}{\pi^{2}}\Sigma_{k=1}^{\infty}(-1)^{k-1}\{\frac{\lambda_{q}^{k}e^{\frac{k\mu_{q}}{T}}+\lambda_{\overline{q}}^{k}e^{-\frac{k\mu_{q}}{T}}}{k^{4}}\}+}\end{array}}\end{array}$}\\
\textcolor{black}{$\begin{array}{c}
\qquad\qquad\qquad\begin{array}{c}
\begin{array}{c}
{\displaystyle \frac{6m_{0s}^{4}}{\pi^{2}}\Sigma_{k=1}^{\infty}(-1)^{k-1}\lambda_{s}^{k}\{\frac{3K_{2}(kx_{s})}{(kx_{s})^{2}}+\frac{K_{1}(kx_{s})}{(kx_{s})}\}\}}\end{array}\end{array}\end{array}$}\\
\textcolor{black}{and }\\
\textcolor{black}{$\begin{array}{c}
{\displaystyle {\displaystyle \begin{array}{c}
{\displaystyle p=\{\frac{16T^{4}}{\pi^{2}}\Sigma_{k=1}^{\infty}\frac{\lambda_{g}^{k}}{k^{4}}+\frac{12T^{4}}{\pi^{2}}\Sigma_{k=1}^{\infty}(-1)^{k-1}\{\frac{\lambda_{q}^{k}e^{\frac{k\mu_{q}}{T}}+\lambda_{\overline{q}}^{k}e^{-\frac{k\mu_{q}}{T}}}{k^{4}}\}+}\end{array}}}\end{array}$}\\
\textcolor{black}{$\begin{array}{c}
{\displaystyle \begin{array}{c}
\qquad\qquad\qquad\begin{array}{c}
\begin{array}{c}
{\displaystyle \frac{6m_{0s}^{4}}{\pi^{2}}\Sigma_{k=1}^{\infty}(-1)^{k-1}\lambda_{s}^{k}\frac{K_{2}(kx_{s})}{(kx_{s})^{2}}\}}\end{array}\end{array}\end{array}}\end{array}$}\\
$.\qquad\qquad\qquad\qquad\qquad\qquad\qquad\qquad\qquad\qquad\qquad\qquad\qquad\qquad\qquad\qquad$...(4)\textcolor{black}{}\\
\textcolor{black}{}\\

\subsection{\textcolor{black}{\normalsize Baryon Number Conservation Equation}\protect \\
\protect \\
}

\textcolor{black}{The net baryon number of the system equals}\\
\textcolor{black}{$\begin{array}{c}
{\displaystyle n_{B}V=\frac{1}{3}(n_{q}-n_{\overline{q}})V}\end{array}$}\\
\textcolor{black}{If we restrict our discussions to a unit volume
then the baryon density equals }\\
\textcolor{black}{$\begin{array}{c}
n_{B}=\frac{1}{3}(n_{q}-n_{\overline{q}})=\frac{4T^{3}}{\pi^{2}}\Sigma_{k=1}^{\infty}(-1)^{k-1}\{\frac{\lambda_{q}^{k}e^{\frac{k\mu_{q}}{T}}-\lambda_{\overline{q}}^{k}e^{-\frac{k\mu_{q}}{T}}}{k^{3}}\}\end{array}$}\\
\textcolor{black}{for given values of the parameters. Thus, if the
initial baryon number density, temperature and non-equilibrium fugacities
be known ( as from models like HIJING or SSPC or likewise ) then we
can iterate to get initial value of the quark chemical potential.
}\\
\textcolor{black}{The baryon number conservation equation gives }\\
\textcolor{black}{$\begin{array}{c}
\partial_{\mu}\end{array}(n_{B}u^{\mu})=\frac{\partial n_{B}}{\partial\tau}+\frac{n_{B}}{\tau}=0$}$\qquad\qquad\qquad\qquad\qquad\qquad\qquad\qquad\qquad$...(5)\textcolor{black}{}\\
\textcolor{black}{from which we can obtain expression for $\begin{array}{c}
\dot{\mu_{q}}\end{array}$, which gives the rate of change of the chemical potential with time.
If we substitute for the number densities and solve for $\begin{array}{c}
{\displaystyle \begin{array}{c}
\dot{\mu_{q}}\end{array}}\end{array}$we get }\\
\textcolor{black}{$\begin{array}{c}
{\displaystyle {\displaystyle \begin{array}{c}
\dot{\mu_{q}}\end{array}}=\dot{\lambda_{q}B_{1}}+}\dot{\dot{\lambda_{\overline{q}}}B_{2}+\dot{T}B_{3}+B_{4}}\end{array}$}$\qquad\qquad\qquad\qquad\qquad$...(6)\textcolor{black}{}\\
\textcolor{black}{where }\\
\textcolor{black}{$\begin{array}{c}
{\displaystyle B_{1}=-\frac{T}{\bigtriangleup}}\sum_{k=1}^{\infty}\frac{(-1)^{k-1}}{k^{2}}\lambda_{q}^{k-1}e^{\frac{k\mu_{q}}{T}}\end{array}$}\\
\textcolor{black}{$\begin{array}{c}
{\displaystyle \begin{array}{c}
{\displaystyle B_{2}=\frac{T}{\bigtriangleup}}\sum_{k=1}^{\infty}\frac{(-1)^{k-1}}{k^{2}}\lambda_{\bar{q}}^{k-1}e^{-\frac{k\mu_{q}}{T}}\end{array}}\end{array}$}\\
\textcolor{black}{$\begin{array}{c}
{\displaystyle B_{3}=-\frac{3}{\Delta}\Sigma_{k=1}^{\infty}(-1)^{k-1}\{\frac{\lambda_{q}^{k}e^{\frac{k\mu_{q}}{T}}+\lambda_{\overline{q}}^{k}e^{-\frac{k\mu_{q}}{T}}}{k^{3}}\}}\end{array}-{\displaystyle \frac{\mu_{q}}{T}}$}\\
\textcolor{black}{and}\\
\textcolor{black}{$\begin{array}{c}
B_{4}\end{array}{\displaystyle =-\frac{{\displaystyle T}}{\tau\bigtriangleup}}\Sigma_{k=1}^{\infty}(-1)^{k-1}\{\frac{\lambda_{q}^{k}e^{\frac{k\mu_{q}}{T}}-\lambda_{\overline{q}}^{k}e^{-\frac{k\mu_{q}}{T}}}{k^{3}}\}$}\\
\textcolor{black}{with}\\
\textcolor{black}{$\begin{array}{c}
\bigtriangleup=\end{array}\Sigma_{k=1}^{\infty}(-1)^{k-1}\{\frac{\lambda_{q}^{k}e^{\frac{k\mu_{q}}{T}}+\lambda_{\overline{q}}^{k}e^{-\frac{k\mu_{q}}{T}}}{k^{2}}\}$}\\
\textcolor{black}{We shall see the effect of having these nonzero
values very shortly.}\\
\textcolor{black}{}\\
\textcolor{black}{}\\

\subsection{\textcolor{black}{\normalsize Quark \& Anti-Quark Number Density
Evolution Equation}\protect \\
\textcolor{black}{\normalsize }\protect \\
}

\textcolor{black}{Using standard techniques we can obtain the fermionic
number density evolution equations. The marked difference from the
zero chemical potential case is that due to the nonzero values of
the coefficients of the Baryon Number Conservation equation, the massless
quark and antiquark number density evolution equations both depend
on either of the two growth rates. Let us put forward the two Number
Density Evolution Equations. We include a new process called the quark
flavour changing process (qfcp) into our framework.}{\large }\\
{\large \par}

\subsubsection{Massless Quark Number Density Evolution Equation\protect \\
}

\textcolor{black}{If the Quark Number Density Evolution Equation be
explicitly worked out it takes the form }\\
\textcolor{black}{$\begin{array}{c}
{\displaystyle {\displaystyle \dot{\dot{\lambda_{q}}Q_{1}}+}\dot{\dot{\lambda_{\overline{q}}}Q_{2}+\dot{T}Q_{3}+Q_{4}=0}}\end{array}$}$\qquad\qquad\qquad\qquad\qquad$...(7)\textcolor{black}{}\\
\textcolor{black}{where}\\
\textcolor{black}{$\begin{array}{c}
{\displaystyle Q_{1}=\frac{12T^{3}}{\pi^{2}}\Sigma_{k=1}^{\infty}(-1)^{k-1}\{\frac{\lambda_{q}^{k-1}e^{\frac{k\mu_{q}}{T}}+\lambda_{q}^{k}e^{\frac{k\mu_{q}}{T}}\frac{B_{1}}{T}}{k^{2}}\}}\end{array}$}\\
\textcolor{black}{$\begin{array}{c}
{\displaystyle \begin{array}{c}
{\displaystyle Q_{2}=\frac{12T^{3}}{\pi^{2}}\Sigma_{k=1}^{\infty}(-1)^{k-1}\frac{\lambda_{q}^{k}e^{\frac{k\mu_{q}}{T}}\frac{B_{2}}{T}}{k^{2}}}\end{array}}\end{array}$}\\
\textcolor{black}{$\begin{array}{c}
{\displaystyle Q_{3}=\frac{36T^{2}}{\pi^{2}}\Sigma_{k=1}^{\infty}(-1)^{k-1}\frac{\lambda_{q}^{k-1}e^{\frac{k\mu_{q}}{T}}}{k^{3}}}+{\displaystyle \frac{12T^{3}}{\pi^{2}}\Sigma_{k=1}^{\infty}(-1)^{k-1}\frac{\lambda_{q}^{k}e^{\frac{k\mu_{q}}{T}}\frac{B_{3}}{T}}{k^{2}}}\end{array}$}\\
\textcolor{black}{and}\\
\textcolor{black}{${\displaystyle \begin{array}{c}
Q_{4}={\displaystyle \frac{12T^{3}}{{\tau\pi}^{2}}\Sigma_{k=1}^{\infty}(-1)^{k-1}\frac{\lambda_{q}^{k}e^{\frac{k\mu_{q}}{T}}}{k^{3}}}+{\displaystyle \frac{12T^{3}}{\pi^{2}}\Sigma_{k=1}^{\infty}(-1)^{k-1}\frac{\lambda_{q}^{k}e^{\frac{k\mu_{q}}{T}}\frac{B_{4}}{T}}{k^{2}}}\end{array}}-SQ$}\\
\textcolor{black}{}\\
\textcolor{black}{where }\\
\textcolor{black}{}\\
\textcolor{black}{$\begin{array}{c}
SQ=(R_{gg\rightarrow q\overline{q}}-R_{q\overline{q}\rightarrow gg})+(R_{s\overline{s}\rightarrow q\overline{q}}-R_{q\overline{q}\rightarrow s\overline{s}})\end{array}$}\\

\subsubsection{\textcolor{black}{Massless Anti-Quark Number Density Evolution Equation}\protect \\
\textcolor{black}{}\protect \\
\textcolor{black}{}\protect \\
}

\textcolor{black}{For anti-quarks the equations are more or less identical
as above with $\begin{array}{c}
{\displaystyle \lambda_{q}\rightarrow\lambda_{\bar{q}}}\end{array}$and $\begin{array}{c}
\mu_{q}\rightarrow-\mu_{q}\end{array}$, except for the modifications in the contributions from $\begin{array}{c}
B_{i}\end{array}$. If the Anti-Quark Number Density Evolution Equation be explicitly
worked out it takes the form }\\
\textcolor{black}{$\begin{array}{c}
{\displaystyle {\displaystyle \dot{\dot{\lambda_{q}}AQ_{1}}+}\dot{\dot{\lambda_{\overline{q}}}AQ_{2}+\dot{T}AQ_{3}+AQ_{4}=0}}\end{array}$} $\qquad\qquad\qquad\qquad\qquad$...(8)\textcolor{black}{}\\
\textcolor{black}{where}\\
\textcolor{black}{$\begin{array}{c}
\begin{array}{c}
{\displaystyle AQ_{1}=-\frac{12T^{3}}{\pi^{2}}\Sigma_{k=1}^{\infty}(-1)^{k-1}\frac{\lambda_{\bar{q}}^{k}e^{-\frac{k\mu_{q}}{T}}\frac{B_{1}}{T}}{k^{2}}}\end{array}\\
AQ_{2}=\frac{12T^{3}}{\pi^{2}}\Sigma_{k=1}^{\infty}(-1)^{k-1}\{\frac{\lambda_{\bar{q}}^{k-1}e^{-\frac{k\mu_{q}}{T}}-\lambda_{\bar{q}}^{k}e^{-\frac{k\mu_{q}}{T}}\frac{B_{2}}{T}}{k^{2}}\}\\
\begin{array}{c}
{\displaystyle AQ_{3}=\frac{36T^{2}}{\pi^{2}}\Sigma_{k=1}^{\infty}(-1)^{k-1}\frac{\lambda_{\bar{q}}^{k-1}e^{-\frac{k\mu_{q}}{T}}}{k^{3}}}-{\displaystyle \frac{12T^{3}}{\pi^{2}}\Sigma_{k=1}^{\infty}(-1)^{k-1}\frac{\lambda_{\bar{q}}^{k}e^{-\frac{k\mu_{q}}{T}}\frac{B_{3}}{T}}{k^{2}}}\end{array}\\
{\displaystyle AQ_{4}={\displaystyle \frac{12T^{3}}{{\tau\pi}^{2}}\Sigma_{k=1}^{\infty}(-1)^{k-1}\frac{\lambda_{\bar{q}}^{k}e^{-\frac{k\mu_{q}}{T}}}{k^{3}}}-{\displaystyle \frac{12T^{3}}{\pi^{2}}\Sigma_{k=1}^{\infty}(-1)^{k-1}\frac{\lambda_{\bar{q}}^{k}e^{-\frac{k\mu_{q}}{T}}\frac{B_{4}}{T}}{k^{2}}-SaQ}}\end{array}$}\\
\textcolor{black}{}\\
\textcolor{black}{where }\\
\textcolor{black}{}\\
\textcolor{black}{$\begin{array}{c}
SaQ=(R_{gg\rightarrow q\overline{q}}-R_{q\overline{q}\rightarrow gg})+(R_{s\overline{s}\rightarrow q\overline{q}}-R_{q\overline{q}\rightarrow s\overline{s}})=SQ\end{array}$}

\subsubsection{Massive Strange Quark \textcolor{black}{Number Density Evolution
Equation}\protect \\
}

For massive strange quark the evolution equation is given by \\
\\
${\displaystyle \dot{\lambda_{s}}S_{1}+\dot{T}S_{2}+S_{3}=0}$$\qquad\qquad\qquad\qquad\qquad$...(9)\\
where\\
$\begin{array}{c}
{\displaystyle S_{1}=\frac{\sum_{k=1}^{\infty}(-1)^{k-1}k\dot{\lambda_{s}^{k-1}{\displaystyle \frac{K_{2}(kx_{s})}{(kx_{s})}}}}{\sum_{k=1}^{\infty}(-1)^{k-1}\dot{\lambda_{s}^{k}{\displaystyle \frac{K_{2}(kx_{s})}{(kx_{s})}}}}}\end{array}$ \\
${\displaystyle S_{2}=}{\displaystyle \frac{3}{T}}{\displaystyle \frac{\sum_{k=1}^{\infty}(-1)^{k-1}\dot{\lambda_{s}^{k}\{{\displaystyle \frac{K_{2}(kx_{s})}{(kx_{s})}+\frac{1}{3}K_{1}(kx_{s})\}}}}{\sum_{k=1}^{\infty}(-1)^{k-1}\dot{\lambda_{s}^{k}{\displaystyle \frac{K_{2}(kx_{s})}{(kx_{s})}}}}}$
\\
${\displaystyle \begin{array}{c}
{\displaystyle S_{3}=}{\displaystyle \frac{1}{\tau}-SQs}\end{array}}$........ ( 11c ) with \\
$\begin{array}{c}
\begin{array}{c}
SQs=\end{array}\{(R_{gg\rightarrow s\overline{s}}-R_{s\overline{s}\rightarrow gg})/n_{s}\}+2\{(R_{q\overline{q}\rightarrow s\overline{s}}-R_{s\overline{s}\rightarrow q\overline{q}})/n_{s}\}\end{array}$\\
{\large }\\
{\large \par}

\subsection{\textcolor{black}{\normalsize Energy-Momentum Conservation Equation}\protect \\
}

\textcolor{black}{The conservation of the energy-momentum tensor ,
which we term as Bjorken's equation for future references,gives }\\
\textcolor{black}{${\displaystyle \frac{d\varepsilon}{d\tau}+\frac{\varepsilon+p}{\tau}=0}$}$\qquad\qquad\qquad\qquad\qquad\qquad\qquad\qquad\qquad$...(10)\textcolor{black}{}\\
\textcolor{black}{Inserting the full expressions of energy density
and pressure we obtain }\\
\textcolor{black}{$\begin{array}{c}
\dot{T}f_{6}+\dot{\lambda_{g}}f_{7}+\dot{\lambda_{q}}f_{8}+\dot{\lambda_{\overline{q}}}f_{9}+\dot{\lambda_{s}}f_{s}+f_{10}=0\end{array}$} $\qquad\qquad\qquad\qquad\qquad$...(11)\textcolor{black}{}\\
\textcolor{black}{where }\\
\textcolor{black}{$\begin{array}{c}
{\displaystyle f_{6}=\frac{192T^{3}}{\pi^{2}}\sum_{k=1}^{\infty}\frac{\lambda_{g}^{k}}{k^{4}}+\frac{36T^{4}}{\pi^{2}}(\frac{B_{3}}{T}-\frac{\mu_{q}}{T^{2}})\Sigma_{k=1}^{\infty}(-1)^{k-1}\{\frac{\lambda_{q}^{k}e^{\frac{k\mu_{q}}{T}}-\lambda_{\overline{q}}^{k}e^{-\frac{k\mu_{q}}{T}}}{k^{3}}\}+}\\
{\displaystyle \frac{144T^{3}}{\pi^{2}}\Sigma_{k=1}^{\infty}(-1)^{k-1}\{\frac{\lambda_{q}^{k}e^{\frac{k\mu_{q}}{T}}+\lambda_{\overline{q}}^{k}e^{-\frac{k\mu_{q}}{T}}}{k^{4}}\}+\frac{72m_{0s}^{4}}{\pi^{2}T}\Sigma_{k=1}^{\infty}(-1)^{k-1}\lambda_{s}^{k}\{\frac{K_{2}(kx_{s})}{(kx_{s})}+}\\
{\displaystyle +\frac{5K_{1}(kx_{s})}{12(kx_{s})}+\frac{K_{0}(kx_{s})}{12}\}}\\
{\displaystyle f_{7}=\frac{48T^{4}}{\pi^{2}}\sum_{k=1}^{\infty}\frac{\lambda_{g}^{k-1}}{k^{3}}}\\
\begin{array}{c}
{\displaystyle f_{8}=\frac{18T^{4}}{\pi^{2}}\Sigma_{k=1}^{\infty}(-1)^{k-1}\{\frac{\lambda_{q}^{k-1}e^{\frac{k\mu_{q}}{T}}}{k^{2}}\}+\frac{18T^{4}}{\pi^{2}}\Sigma_{k=1}^{\infty}(-1)^{k-1}\frac{(\lambda_{q}^{k-1}e^{\frac{k\mu_{q}}{T}}-\lambda_{q}^{k}e^{\frac{k\mu_{q}}{T}})\frac{B_{1}}{T}}{k^{3}}}\end{array}\\
\begin{array}{c}
{\displaystyle f_{9}=\frac{18T^{4}}{\pi^{2}}\Sigma_{k=1}^{\infty}(-1)^{k-1}\{\frac{\lambda_{\bar{q}}^{k-1}e^{-\frac{k\mu_{q}}{T}}}{k^{2}}\}+\frac{18T^{4}}{\pi^{2}}\Sigma_{k=1}^{\infty}(-1)^{k-1}\frac{(\lambda_{q}^{k-1}e^{\frac{k\mu_{q}}{T}}-\lambda_{q}^{k}e^{\frac{k\mu_{q}}{T}})\frac{B_{2}}{T}}{k^{3}}}\end{array}\\
{\displaystyle f_{s}=\frac{6m_{0s}^{4}}{\pi^{2}}\Sigma_{k=1}^{\infty}(-1)^{k-1}k.\lambda_{s}^{k-1}\{\frac{3K_{2}(kx_{s})}{(kx_{s})}+\frac{K_{1}(kx_{s})}{(kx_{s})}}\}\\
\begin{array}{c}
{\displaystyle f_{10}=\frac{64T^{4}}{\tau\pi^{2}}\sum_{k=1}^{\infty}\frac{\lambda_{g}^{k}}{k^{4}}+\frac{48T^{4}}{\tau\pi^{2}}\Sigma_{k=1}^{\infty}(-1)^{k-1}\{\frac{\lambda_{q}^{k}e^{\frac{k\mu_{q}}{T}}+\lambda_{\overline{q}}^{k}e^{-\frac{k\mu_{q}}{T}}}{k^{4}}\}+}\\
{\displaystyle \frac{6m_{0s}^{4}}{\tau\pi^{2}}\Sigma_{k=1}^{\infty}(-1)^{k-1}\lambda_{s}^{k}\{\frac{4K_{2}(kx_{s})}{(kx_{s})}+}{\displaystyle \frac{K_{1}(kx_{s})}{(kx_{s})}}\}+\\
{\displaystyle \frac{36T^{3}}{\tau\pi^{2}}\Sigma_{k=1}^{\infty}(-1)^{k-1}\{\frac{\lambda_{q}^{k}e^{\frac{k\mu_{q}}{T}}-\lambda_{\overline{q}}^{k}e^{-\frac{k\mu_{q}}{T}}}{k^{3}}\}B_{4}}\end{array}\end{array}$}{\large }\\
{\large \par}

\section{\textcolor{black}{\normalsize Changing Distribution Functions}\protect \\
}

\textcolor{black}{In order to reduce computational complications to
a low degree we make certain approximations.We redefine the non-equilibrium
distribution functions and shift from the Juttner distribution function
to Modified Fermi-Dirac type for the quarks and antiquarks in confirmation
with earlier works {[}4,5]. Needless to say, this would surely limit
the applicability of the present study. However, it is believed that
the basic nature of partonic fugacity variations would remain unaltered.Since
we adopt certain approximations for reasons as mentioned earlier,
we refrain from giving the explicit forms of the reaction rates presently
and hold back the expressions thereof for now. }\\

\textcolor{black}{As stated, we adopt the following approximations:}\\
\textcolor{black}{}\\
\textcolor{black}{We replace the original distribution functions by
the following:}\\
\textcolor{black}{$\begin{array}{c}
{\displaystyle {\displaystyle f_{g}=\frac{\lambda_{g}}{e^{\varepsilon/T}-1}}}\\
{\displaystyle f_{q(\overline{q})}=\frac{\lambda_{q(\overline{q})}e^{\pm\frac{\mu_{q}}{T}}}{e^{\varepsilon/T}+1}}\\
{\displaystyle f_{s}=f_{\overline{s}}=\frac{\lambda_{s}}{e^{\varepsilon/T}+1}}\end{array}$}$\qquad\qquad\qquad\qquad\qquad$...(12)\textcolor{black}{}\\
\textcolor{black}{Of these, the last two are the Modified Fermi-Dirac
( MFD ) type Distribution functions as was used in {[}4,5] . }\\
\textcolor{black}{}\\
\textcolor{black}{As for the quarks, we take }\\
\textcolor{black}{i) For s quark we take mass\textasciitilde{} 150
MeV. }\\
\textcolor{black}{ii) For lighter quarks since mass is much too less
than the strange quark mass, we treat the mass of lighter flavours
as zero.}\\
\textcolor{black}{iii) Since the lighter quarks are treated to have
the same mass, we neglect $\begin{array}{c}
{\displaystyle u\overline{u}\rightleftharpoons d\overline{d}}\end{array}$type of reactions as they would be like elastic scattering processes
leading to no net particle production.}

\subsection{\textcolor{black}{\normalsize Basic Thermodynamic Quantities}\protect \\
\textcolor{black}{\normalsize }\protect \\
}

\textcolor{black}{Using the techniques of statistical mechanics, we
can obtain the number density, energy density and pressure as follows:
}\\
\textcolor{black}{}\\
$\begin{array}{c}
{\displaystyle n=\frac{16\zeta(3)T^{3}}{\pi^{2}}\lambda_{g}+\frac{9\zeta(3)n_{f}T^{3}}{2\pi^{2}}\{\lambda_{q}e^{\mu_{q}/T}+\lambda_{\overline{q}}e^{{-\mu}_{q}/T}\}+}\end{array}$\\
$\begin{array}{c}
{\displaystyle \frac{6m_{0s}^{3}}{\pi^{2}}\lambda_{s}\sum_{k=1}^{\infty}(-1)^{k-1}\frac{K_{2}(kx_{s})}{(kx_{s})}}\end{array}$\\
$\begin{array}{c}
{\displaystyle {\displaystyle {\displaystyle \varepsilon=\frac{8\pi^{2}T^{4}}{15}\lambda_{g}+\frac{7\pi^{2}n_{f}T^{4}}{40}\{\lambda_{q}e^{\mu_{q}/T}+\lambda_{\overline{q}}e^{{-\mu}_{q}/T}\}+}}}\end{array}$\\
$\begin{array}{c}
{\displaystyle \begin{array}{c}
{\displaystyle \frac{6m_{0s}^{4}}{\pi^{2}}\lambda_{s}\sum_{k=1}^{\infty}(-1)^{k-1}\{\frac{{3K}_{2}(kx_{s})}{(kx_{s})^{2}}+}{\displaystyle \frac{K_{1}(kx_{s})}{(kx_{s})}\}}\end{array}}\end{array}$\\
$\begin{array}{c}
{\displaystyle p=\frac{8\pi^{2}T^{4}}{45}\lambda_{g}+\frac{7\pi^{2}n_{f}T^{4}}{120}\{\lambda_{q}e^{\mu_{q}/T}+\lambda_{\overline{q}}e^{{-\mu}_{q}/T}\}+}\end{array}$\\
$\begin{array}{c}
{\displaystyle {\displaystyle \begin{array}{c}
{\displaystyle \frac{6m_{0s}^{4}}{\pi^{2}}\lambda_{s}\sum_{k=1}^{\infty}(-1)^{k-1}\frac{K_{2}(kx_{s})}{(kx_{s})^{2}}}\end{array}}}\end{array}$\textcolor{black}{}\\
$with\begin{array}{c}
\, x_{s}{=m}_{s}/T\end{array}$\textcolor{black}{}\\
\textcolor{black}{$\begin{array}{c}
\qquad\qquad\qquad\qquad\qquad\qquad\qquad\qquad\qquad\qquad\qquad...(13)\end{array}$}\\

\textcolor{black}{Using standard techniques we can obtain the quark
and anti-quark number density evolution equations. As for the Juttner
distribution case, the equations are coupled via the Baryon Number
Conservation Equation coefficients $\begin{array}{c}
{\displaystyle B_{i}}\end{array}$.}\\
\textcolor{black}{}\\
\textcolor{black}{}\\

\subsection{\textcolor{black}{\normalsize Baryon Number Conservation Equation}\protect \\
\textcolor{black}{\normalsize }\protect \\
\textcolor{black}{\normalsize }\protect \\
}

\textcolor{black}{Using the modified distribution functions we can
obtain the coefficients $\begin{array}{c}
B_{i}\end{array}$. We obtain }\\
\textcolor{black}{}\\
\textcolor{black}{${\displaystyle {\displaystyle {\displaystyle \begin{array}{c}
\dot{\mu_{q}}\end{array}}=\dot{\dot{\lambda_{q}}B_{1}}+}\dot{\dot{\lambda_{\overline{q}}}B_{2}+\dot{T}B_{3}+B_{4}}}$}$\qquad\qquad\qquad\qquad\qquad$...(14)\textcolor{black}{}\\
\textcolor{black}{where}\\
\textcolor{black}{$\begin{array}{c}
{\displaystyle B_{1}=-\frac{Te^{\frac{\mu_{q}}{T}}}{\Delta_{1}}}\end{array}$}\\
\textcolor{black}{$\begin{array}{c}
{\displaystyle B_{2}=\frac{Te^{-\frac{\mu_{q}}{T}}}{\Delta_{1}}}\end{array}$}\\
\textcolor{black}{$\begin{array}{c}
{\displaystyle {\displaystyle B_{3}=\frac{\mu_{q}}{T}-\frac{\Delta_{2}}{\Delta_{1}}}}\end{array}$}\\
\textcolor{black}{$\begin{array}{c}
{\displaystyle B_{4}=-\frac{T}{\tau}\frac{\Delta_{2}}{\Delta_{1}}}\end{array}$}\\
\textcolor{black}{where}\\
\textcolor{black}{$\begin{array}{c}
{\displaystyle \Delta_{1}=\lambda_{i}e^{\frac{\mu_{i}}{T}}+\lambda_{\overline{i}}e^{-\frac{\mu_{i}}{T}}}\\
\Delta_{2}=\lambda_{i}e^{\frac{\mu_{i}}{T}}-\lambda_{\overline{i}}e^{-\frac{\mu_{i}}{T}}\end{array}$}

\subsection{\textcolor{black}{\normalsize Quark and Anti-Quark Number Density
Evolution Equations}\protect \\
\textcolor{black}{\normalsize }\protect \\
}

\textcolor{black}{Just as in the case of the full Juttner distribution,
here too the quark and anti-quark number density evolution equations
are coupled via the coefficients $\begin{array}{c}
{\displaystyle B_{i}}\end{array}$. }\\
\textcolor{black}{}\\

\subsubsection{\textcolor{black}{Massless Quark Number Density Evolution Equation}\protect \\
\textcolor{black}{}\protect \\
}

\textcolor{black}{The equation is given by }\\
\textcolor{black}{}\\
\textcolor{black}{${\displaystyle {\displaystyle \dot{\dot{\lambda_{q}}Q_{1}}+}\dot{\dot{\lambda_{\overline{q}}}Q_{2}+\dot{T}Q_{3}+Q_{4}=0}}$}$\qquad\qquad\qquad\qquad\qquad$...(15)\textcolor{black}{}\\
\textcolor{black}{where}\\
\textcolor{black}{$\begin{array}{c}
{\displaystyle Q_{1}=\frac{1}{\lambda_{q}}+\frac{B_{1}}{T}}\end{array}$}\\
\textcolor{black}{$\begin{array}{c}
\begin{array}{c}
{\displaystyle Q_{2}=\frac{B_{2}}{T}}\end{array}\end{array}$}\\
\textcolor{black}{$\begin{array}{c}
{\displaystyle Q_{3}=\frac{3}{T}}\end{array}{\displaystyle +\frac{B_{3}}{T}-\frac{\mu_{q}}{T^{2}}}$}\\
\textcolor{black}{${\displaystyle \begin{array}{c}
{\displaystyle Q_{4}=}{\displaystyle \frac{1}{\tau}+\frac{B_{4}}{T}}\end{array}-SQ}$ with}\\
\textcolor{black}{$\begin{array}{c}
\begin{array}{c}
\begin{array}{c}
SQ=\end{array}\{(R_{gg\rightarrow q\overline{q}}-R_{q\overline{q}\rightarrow gg})/n_{q}\}-\{(R_{q\overline{q}\rightarrow s\overline{s}}-R_{s\overline{s}\rightarrow q\overline{q}})/n_{q}\}\end{array}\end{array}$}\\
{\large }\\
{\large \par}

\subsubsection{Massless Anti-Quark Number Density Evolution Equation\protect \\
\protect \\
}

\textcolor{black}{The equation is given by}\\
\textcolor{black}{${\displaystyle {\displaystyle \dot{\dot{\lambda_{q}}Q_{1}}+}\dot{\dot{\lambda_{\overline{q}}}Q_{2}+\dot{T}Q_{3}+Q_{4}=0}}$}$\qquad\qquad\qquad\qquad\qquad$...(16)\textcolor{black}{}\\

\textcolor{black}{${\displaystyle \begin{array}{c}
{\displaystyle {\displaystyle AQ_{1}=-\frac{B_{1}}{T}}}\end{array}}$}\\
\textcolor{black}{${\displaystyle \begin{array}{c}
{\displaystyle \begin{array}{c}
{\displaystyle AQ_{2}=\frac{1}{\lambda_{\overline{q}}}-\frac{B_{2}}{T}}\end{array}}\end{array}}$}\\
\textcolor{black}{$\begin{array}{c}
{\displaystyle \begin{array}{c}
{\displaystyle AQ_{3}=\frac{3}{T}}\end{array}{\displaystyle -\frac{B_{3}}{T}+\frac{\mu_{q}}{T^{2}}}}\end{array}$}\\
\textcolor{black}{$\begin{array}{c}
{\displaystyle {\displaystyle \begin{array}{c}
{\displaystyle AQ_{4}=}{\displaystyle \frac{1}{\tau}-\frac{B_{4}}{T}}-SaQ\end{array}}}\end{array}$with}\\
$\begin{array}{c}
\begin{array}{c}
SaQ=\end{array}\{(R_{gg\rightarrow q\overline{q}}-R_{q\overline{q}\rightarrow gg})/n_{\overline{q}}\}-\{(R_{q\overline{q}\rightarrow s\overline{s}}-R_{s\overline{s}\rightarrow q\overline{q}})/n_{\overline{q}}\}\end{array}$\\
\\

\subsubsection{Massive Strange Quark \textcolor{black}{Number Density Evolution
Equation}\protect \\
}

For massive strange quark the evolution equation is given by \\
\\
${\displaystyle \dot{\lambda_{s}}S_{1}+\dot{T}S_{2}+S_{3}=0}$$\qquad\qquad\qquad\qquad\qquad$...(17)\\
where\\
$\begin{array}{c}
{\displaystyle S_{1}=\frac{1}{\dot{\lambda_{s}}}}\end{array}$ \\
${\displaystyle S_{2}=}{\displaystyle \frac{3}{T}}{\displaystyle \frac{\sum_{k=1}^{\infty}(-1)^{k-1}\dot{\{{\displaystyle \frac{K_{2}(kx_{s})}{(kx_{s})}+\frac{1}{3}K_{1}(kx_{s})\}}}}{\sum_{k=1}^{\infty}(-1)^{k-1}\dot{\lambda_{s}^{k}{\displaystyle \frac{K_{2}(kx_{s})}{(kx_{s})}}}}}$
\\
${\displaystyle \begin{array}{c}
{\displaystyle S_{3}=}{\displaystyle \frac{1}{\tau}-SQs}\end{array}}$with \\
$\begin{array}{c}
\begin{array}{c}
SQs=\end{array}\{(R_{gg\rightarrow s\overline{s}}-R_{s\overline{s}\rightarrow gg})/n_{s}\}+2\{(R_{q\overline{q}\rightarrow s\overline{s}}-R_{s\overline{s}\rightarrow q\overline{q}})/n_{s}\}\end{array}$\\
{\large }\\
\textcolor{black}{}\\

\subsection{\textcolor{black}{\normalsize Bjorken's Equation}\protect \\
\textcolor{black}{\normalsize }\protect \\
}

\textcolor{black}{From Bjorken's equation we get on substituting for
energy and momentum}\\
\textcolor{black}{$\begin{array}{c}
{\displaystyle \begin{array}{c}
\dot{T}f_{6}+\dot{\lambda_{g}}f_{7}+\dot{\lambda_{q}}f_{8}+\dot{\lambda_{\overline{q}}}f_{9}+\dot{\lambda_{s}}f_{s}+f_{10}=0\end{array}}\end{array}$}$\qquad\qquad\qquad$...(18)\textcolor{black}{}\\
\textcolor{black}{}\\
\textcolor{black}{where}\\
\textcolor{black}{}\\

\textcolor{black}{${\displaystyle \begin{array}{c}
{\displaystyle f_{6}=\frac{32\pi^{2}}{15}\lambda_{g}T^{3}+}\end{array}}$}\\
\textcolor{black}{$\begin{array}{c}
{\displaystyle \frac{7{\displaystyle \pi^{2}n_{f}}}{40}T^{4}(\lambda_{q}e^{\frac{\mu_{q}}{T}}-\lambda_{\overline{q}}e^{-\frac{\mu_{q}}{T}})(\frac{B_{3}}{T}-\frac{\mu_{q}}{T^{2}})+\frac{28{\displaystyle \pi^{2}n_{f}}}{40}T^{3}(\lambda_{q}e^{\frac{\mu_{q}}{T}}+\lambda_{\overline{q}}e^{-\frac{\mu_{q}}{T}})+}\end{array}$}\\
\textcolor{black}{$\begin{array}{c}
{\displaystyle {\displaystyle \frac{72m_{0s}^{4}}{\pi^{2}T}\lambda_{s}\sum_{k=1}^{infty}(-1)^{k-1}}\{\frac{K_{2}(kx_{s})}{(kx_{s})^{2}}}{\displaystyle +\frac{5K_{1}(kx_{s})}{12(kx_{s})}+\frac{K_{0}(kx_{s})}{12}\}}\end{array}$}\\
\textcolor{black}{$\begin{array}{c}
{\displaystyle f_{7}=\frac{8\pi^{2}}{15}T^{4}}\end{array}$}\\
\textcolor{black}{$\begin{array}{c}
f_{s}=\end{array}\begin{array}{c}
{\displaystyle \frac{6m_{0s}^{4}}{\pi^{2}}\sum_{k=1}^{\infty}(-1)^{k-1}\{\frac{{3K}_{2}(kx_{s})}{(kx_{s})^{2}}+\frac{K_{1}(kx_{s})}{(kx_{s})}\}}\end{array}$}\\
\textcolor{black}{$\begin{array}{c}
{\displaystyle f_{8}=\frac{7{\displaystyle \pi^{2}n_{f}}}{40}\{{\displaystyle e^{\frac{\mu_{q}}{T}}}+(\lambda_{q}e^{\frac{\mu_{q}}{T}}-\lambda_{\overline{q}}e^{-\frac{\mu_{q}}{T}})\frac{B_{1}}{T}}\}\end{array}$}\\
\textcolor{black}{$\begin{array}{c}
{\displaystyle {\displaystyle f_{9}=\frac{7{\displaystyle \pi^{2}n_{f}}}{40}\{{\displaystyle e^{-\frac{\mu_{q}}{T}}}+(\lambda_{q}e^{\frac{\mu_{q}}{T}}-\lambda_{\overline{q}}e^{-\frac{\mu_{q}}{T}})\frac{B_{2}}{T}}\}}\end{array}$}\\
\textcolor{black}{$\begin{array}{c}
{\displaystyle f_{10}=\frac{32\pi^{2}}{45\tau}\lambda_{g}T^{4}+\frac{6m_{0s}^{4}}{\pi^{2}\tau}\lambda_{s}\Sigma_{k=1}^{\infty}(-1)^{k-1}\{\frac{4K_{2}(kx_{i})}{(kx_{i})^{2}}+\frac{K_{1}(kx_{i})}{(kx_{i})}\}}\end{array}$}\\
\textcolor{black}{$\begin{array}{c}
{\displaystyle +\frac{28{\displaystyle \pi^{2}n_{f}}}{120}T^{4}(\lambda_{q}e^{\frac{\mu_{q}}{T}}+\lambda_{\overline{q}}e^{-\frac{\mu_{qi}}{T}})+\frac{7{\displaystyle \pi^{2}n_{f}}}{40}T^{4}(\lambda_{q}e^{\frac{\mu_{q}}{T}}-\lambda_{\overline{q}}e^{-\frac{\mu_{q}}{T}})\frac{B_{4}}{T}}\end{array}$}\\
\textcolor{black}{}\\
\textcolor{black}{Now,it is to be noted as the quark and anti-quark
number density evolution equations and the baryon number conservation
equation are mathematically very simply connected, so these two evolution
equations linked by the four B coefficients along with the energy-momentum
conservation equation with inputs from the gluon and strange quark
number density evolution equations do not constitute a set of solvable
equations as the system determinant vanishes. To make the system solvable,
we expand the exponential term in the numerators of the first two
B coefficients and thereby introduce some 'assymetry'. Strictly speaking,
this however would mean not exact baryon conservation. }\\

\subsection{\textcolor{black}{\normalsize The Partonic Reaction Rates}}

\subsubsection{\textcolor{black}{The Gluon Multiplication Rate $\protect\begin{array}{c}
R_{3}\protect\end{array}$}}

\textcolor{black}{The gluon multiplication rate has been calculated
by Xiong et. al. {[} 10 ]. The rate of the reaction $\begin{array}{c}
{\displaystyle gg\rightarrow(n-2)g}\end{array}$is given. By explicitly calculating the matrix element {[} 11] (summed
over all the final states and averaged over all initial states) we
can obtain the gluon multiplication rate. }\\
\textcolor{black}{However, to avoid the huge calculations of evaluating
25 Feynman diagrams {[} 11 ] involved, we fall back on previous results
that earlier workers had used . We postulate that the gluon multiplication
rate depends on the chemical potential via the Debye Screening mass.}\\
\textcolor{black}{The Debye Screening mass suitable for a multicomponent
chemically non-equilibrated parton plasma is given by {[} 12 ]}\\
\textcolor{black}{$\begin{array}{c}
{\displaystyle m_{D}^{2}=\frac{2g^{2}}{\pi^{2}}\int dk}k[N_{c}f_{g}+\sum_{i}f_{i}]\end{array}$}$\qquad\qquad\qquad\qquad\qquad\qquad$...(19)\textcolor{black}{}\\
\textcolor{black}{where the sum runs over all flavours i, while $\begin{array}{c}
N_{c}\end{array}$gives the number of colours. To accomodate for antiquarks and remembering
that our number of flavours is 3 and not 6, we propose the following
modification:}\\
\textcolor{black}{$\begin{array}{c}
{\displaystyle \begin{array}{c}
{\displaystyle m_{D}^{2}=\frac{2g^{2}}{\pi^{2}}\int dk}k[3f_{g}+\begin{array}{c}
{\displaystyle \frac{1}{2}}\end{array}\sum_{i=u,d,s}(f_{i}+f_{\overline{i}})]\end{array}}\end{array}$}$\qquad\qquad$...(20)\textcolor{black}{}\\
\textcolor{black}{Using standard techniques , we can obtain the following
results for the mean free path:}\\
\textcolor{black}{$\begin{array}{c}
{\displaystyle \lambda_{f}^{-1}}=n_{g}\int dq_{\bot}^{2}\frac{{\displaystyle d\sigma_{el}^{gg}}}{{\displaystyle d}q_{\bot}^{2}}=\frac{{\displaystyle {\displaystyle 9n_{g}\pi\alpha_{s}^{2}}}}{{\displaystyle 2m_{D}^{2}(1+\frac{2}{9}\frac{m_{D}^{2}}{T^{2}})}}\end{array}$}$\qquad\qquad\qquad\qquad\qquad$...(21)\textcolor{black}{}\\
\textcolor{black}{which for zero chemical potential reduces to the
well known result:}\\
\textcolor{black}{$\begin{array}{c}
{\displaystyle {\displaystyle \lambda_{f}^{-1}}=\frac{9}{8}a_{1}\alpha_{s}T\frac{1}{1+8\pi\alpha_{s}\lambda_{g}/9}}\end{array}$}\\
\textcolor{black}{Using methods of our previous chapter we get on
integrating the modified differential cross section and recalling
the definition of $\begin{array}{c}
{\displaystyle R_{3}=\frac{1}{2}\sigma_{3}n_{g}}\end{array}$the required rate as }\\
\textcolor{black}{$\begin{array}{c}
{\displaystyle \frac{R_{3}}{T}=\frac{27\alpha_{s}^{3}}{2}\lambda_{f}^{2}n_{g}I(\lambda_{g})}\end{array}$}$\qquad\qquad\qquad\qquad\qquad$...(22)\textcolor{black}{}\\
\textcolor{black}{where}\\
\textcolor{black}{$\begin{array}{c}
{\displaystyle I(\lambda_{g})=\int_{1}^{\sqrt{s}\lambda_{f}}dx\int_{0}^{{\displaystyle \frac{s}{4m_{D}^{2}}}}dz\frac{z}{(1+z)^{2}}[\frac{cosh^{-1}\sqrt{x}}{x\sqrt{[x+(1+z)x_{D}]^{2}-4xzx_{D}}}+}\end{array}$}\\
\textcolor{black}{$\begin{array}{c}
{\displaystyle \qquad\qquad\qquad\qquad\qquad\qquad\frac{1}{s\lambda_{f}^{2}}}{\displaystyle \frac{cosh^{-1}\sqrt{x}}{\sqrt{[1+x(1+z)y_{D}]^{2}-4xzy_{D}}}]}\end{array}$}\\
\textcolor{black}{with}\\
\textcolor{black}{$\begin{array}{c}
{\displaystyle x_{D}=m_{D}^{2}\lambda_{f}}\\
{\displaystyle y_{D}=\frac{m_{D}^{2}}{s}}\end{array}$}

\subsubsection{\textcolor{black}{The Quark Anti-Quark pair production rate}}

\textcolor{black}{We have parton production rates in the RHS of the
number density evolution equations.Let us evaluate the quark-antiquark
pair production reaction rate $\begin{array}{c}
R_{2g}\end{array}$. We have {[} 8 ]}\\
\textcolor{black}{$\begin{array}{c}
R_{2g}=R_{{\displaystyle _{gain}}}^{{\displaystyle gg}}-R_{{\displaystyle _{loss}}}^{{\displaystyle gg}}\end{array}$}$\qquad\qquad\qquad\qquad\qquad$...(23)\textcolor{black}{}\\
\textcolor{black}{where}\\
\textcolor{black}{$\begin{array}{c}
{\displaystyle R_{{\displaystyle _{gain}}}^{{\displaystyle gg}}=\int\frac{d^{3}p_{1}}{(2\pi)^{^{{\displaystyle 3}}}2E_{1}}\int\begin{array}{c}
{\displaystyle \frac{d^{3}p_{2}}{(2\pi)^{^{{\displaystyle 3}}}2E_{2}}}\end{array}\int\begin{array}{c}
{\displaystyle \frac{d^{3}p_{3}}{(2\pi)^{^{{\displaystyle 3}}}2E_{3}}}\end{array}\int\begin{array}{c}
\begin{array}{c}
{\displaystyle \frac{d^{3}p_{4}}{(2\pi)^{^{{\displaystyle 3}}}2E_{4}}(2\pi)^{4}}\end{array}\end{array}}\end{array}$}\\
\textcolor{black}{$\begin{array}{c}
{\displaystyle \delta^{4}(p_{1}+p_{2}-p_{3}-p_{4})\Sigma|M_{gg\rightarrow i\overline{i}}|^{{\displaystyle ^{2}}}f_{g}(p_{1})f_{g}(p_{2})(1-f_{q}(p_{3}))(1-f_{\bar{q}}(p_{4}))}\end{array}$}$\qquad$...(23a)\textcolor{black}{}\\
\textcolor{black}{and}\\
\textcolor{black}{$\begin{array}{c}
{\displaystyle R_{{\displaystyle _{loss}}}^{{\displaystyle gg}}=\int\frac{d^{3}p_{1}}{(2\pi)^{^{{\displaystyle 3}}}2E_{1}}\int\begin{array}{c}
{\displaystyle \frac{d^{3}p_{2}}{(2\pi)^{^{{\displaystyle 3}}}2E_{2}}}\end{array}\int\begin{array}{c}
{\displaystyle \frac{d^{3}p_{3}}{(2\pi)^{^{{\displaystyle 3}}}2E_{3}}}\end{array}\int\begin{array}{c}
\begin{array}{c}
{\displaystyle \frac{d^{3}p_{4}}{(2\pi)^{^{{\displaystyle 3}}}2E_{4}}(2\pi)^{4}}\end{array}\end{array}}\end{array}$}\\
\textcolor{black}{$\begin{array}{c}
{\displaystyle {\displaystyle \delta^{4}(p_{1}+p_{2}-p_{3}-p_{4})\Sigma|M_{gg\rightarrow i\overline{i}}|^{{\displaystyle ^{2}}}(1+f_{g}(p_{1}))(1+f_{g}(p_{2}))f_{q}(p_{3})f_{\bar{q}}}}(p_{4})\end{array}$}$\qquad$...(23b)\textcolor{black}{}\\
\textcolor{black}{Following {[} 8 ], we can say that there are three
topologically distinct Feynmann diagrams that contribute towards the
quark-antiquark pair production process. Evaluating them performing
traces and finally adding them up we can find the net squared matrix
element. We basically follow the lines of {[} 8 ]. Transforming variables
as }\\
\textcolor{black}{$\begin{array}{c}
{\displaystyle q=p_{1}+p_{2}}\\
{\displaystyle {\displaystyle p=\frac{1}{2}(p_{1}-p_{2})}}\\
{\displaystyle {\displaystyle q'=p_{3}+p_{4}}}\\
{\displaystyle p'=\frac{1}{2}(p_{3}-p_{4})}\end{array}$}\\
\textcolor{black}{with restrictions}\\
\textcolor{black}{$\begin{array}{c}
{\displaystyle q_{0}>2m_{i}}\\
{\displaystyle s=q_{0}^{2}-|\overrightarrow{\mathbf{q|}}}^{2}\geq4m_{i}^{2}\\
{\displaystyle {\displaystyle p_{0}^{2}\leq\frac{q^{2}}{4}}}\\
{\displaystyle p'_{0}.p'_{0}\leq\frac{q^{2}}{4}(1-\frac{4m_{i}^{2}}{s})}\end{array}$}\\
\textcolor{black}{and transforming the three dimensional integrals
to four dimensional integrals using}\\
\textcolor{black}{${\displaystyle \begin{array}{c}
{\displaystyle \int}\end{array}\frac{d^{3}p_{i}}{2E_{i}}=\int d^{4}p_{i}\delta(p^{2}-m_{i}^{2})}$}\\
\textcolor{black}{with the new set of variables }\\
\textcolor{black}{$\begin{array}{c}
{\displaystyle q_{0}=-Tln}\mathit{v}+2m_{i}\\
{\displaystyle {\displaystyle {\displaystyle q^{\frac{1}{2}}}=(q_{0}^{2}-4m_{i}^{2})^{\frac{1}{2}}u}}\\
{\displaystyle p_{0}=\frac{q}{2}(1-\frac{4m_{i}^{2}}{s})^{\frac{1}{2}}x}\\
{\displaystyle p'_{0}=\frac{q}{2}y}\end{array}$}\\
\textcolor{black}{we arrive at the rate }\\
\textcolor{black}{$\begin{array}{c}
{\displaystyle R_{2g}=\frac{\alpha_{s}^{2}}{2\pi^{3}}T\int_{0}^{1}du\int_{0}^{1}dv\int_{0}^{1}dx\int_{0}^{1}dy\frac{u^{2}}{v}(1-\frac{4m_{i}^{2}}{s})^{\frac{1}{2}}(q_{0}^{2}-4m_{i}^{2})^{\frac{3}{2}}f_{Quarks}f_{phase1}}\end{array}$}...(24)\textcolor{black}{}\\
\textcolor{black}{where}\\
\textcolor{black}{$\begin{array}{c}
{\displaystyle f_{Quarks}=f_{g}(\frac{q_{0}}{2}+p_{0})f_{g}(\frac{q_{0}}{2}-p_{0})(1-f_{q}(\frac{q_{0}}{2}+p'_{0}))(1-f_{\overline{q}}(\frac{q_{0}}{2}-p'_{0}))-}\end{array}$}\\
\textcolor{black}{$\begin{array}{c}
{\displaystyle \qquad\qquad}\end{array}(1+f_{g}(\frac{q_{0}}{2}+p_{0}))(1+f_{g}(\frac{q_{0}}{2}-p_{0}))f_{q}(\frac{q_{0}}{2}+p'_{0})f_{\overline{q}}(\frac{q_{0}}{2}-p'_{0})$}\\
\textcolor{black}{and}\\
\textcolor{black}{$\begin{array}{c}
{\displaystyle f_{phase1}=A+B[\frac{1}{K_{+}}+\frac{1}{K_{-}}]+C[\frac{\bigtriangleup_{+}}{K_{+}^{3}}+\frac{\bigtriangleup_{-}}{K_{+}^{3}}]}\end{array}$}\\
\textcolor{black}{with}\\
\textcolor{black}{}\\
\textcolor{black}{$\begin{array}{c}
{\displaystyle {\displaystyle A=3[1-}[1-{\displaystyle \frac{4m_{i}^{2}}{s}}][{\displaystyle \frac{(1-x^{2})(1-y^{2})}{2}}+x^{2}y^{2}]]-\frac{{\displaystyle 34}}{{\displaystyle 3}}-24{\displaystyle \frac{m_{i}^{2}}{s}}}\end{array}$}\\
\textcolor{black}{$\begin{array}{c}
{\displaystyle {\displaystyle B=\frac{16}{3}[1+\frac{4m_{i}^{2}}{s}+\frac{m_{i}^{4}}{s^{2}}]}}\end{array}$}\\
\textcolor{black}{$\begin{array}{c}
{\displaystyle {\displaystyle C=-\frac{128}{3}{\displaystyle \frac{m_{i}^{4}}{s^{2}}}}}\end{array}$}\\
\textcolor{black}{$\begin{array}{c}
{\displaystyle K_{\pm}={\displaystyle [1-}[1-{\displaystyle \frac{4m_{i}^{2}}{s}}][(1-x^{2}-y^{2})\pm2[1-{\displaystyle \frac{4m_{i}^{2}}{s}}]^{\frac{1}{2}}xy]^{\frac{1}{2}}}\end{array}$}\\
\textcolor{black}{}\\

\subsubsection{\textcolor{black}{The Quark flavour changing rate }}

\textcolor{black}{For the quark flavour changing process we have {[}
8 ]}\\
\textcolor{black}{$\begin{array}{c}
R_{qg}={\displaystyle R^{{\displaystyle q\overline{q}}}}-R_{{\displaystyle _{loss}}}^{{\displaystyle q\overline{q}}}\end{array}$}$\qquad\qquad\qquad\qquad\qquad$...(25)\textcolor{black}{}\\
\textcolor{black}{where}\\
\textcolor{black}{$\begin{array}{c}
{\displaystyle R_{{\displaystyle _{gain}}}^{q\overline{q}}=\int\frac{d^{3}p_{1}}{(2\pi)^{^{{\displaystyle 3}}}2E_{1}}\int\begin{array}{c}
{\displaystyle \frac{d^{3}p_{2}}{(2\pi)^{^{{\displaystyle 3}}}2E_{2}}}\end{array}\int\begin{array}{c}
{\displaystyle \frac{d^{3}p_{3}}{(2\pi)^{^{{\displaystyle 3}}}2E_{3}}}\end{array}\int\begin{array}{c}
\begin{array}{c}
{\displaystyle \frac{d^{3}p_{4}}{(2\pi)^{^{{\displaystyle 3}}}2E_{4}}(2\pi)^{4}}\end{array}\end{array}}\end{array}$}\\
\textcolor{black}{$\begin{array}{c}
{\displaystyle \delta(p_{1}+p_{2}-p_{3}-p_{4})\Sigma|M_{s\overline{s}\rightarrow q\overline{q}}|^{{\displaystyle ^{2}}}f_{q}(p_{1})f_{\overline{q}}(p_{2})(1-f_{s}(p_{3}))(1-f_{\bar{s}}(p_{4}))}\end{array}$}$\qquad$...(25a)\textcolor{black}{}\\
\textcolor{black}{and}\\
\textcolor{black}{$\begin{array}{c}
{\displaystyle R_{{\displaystyle _{loss}}}^{q\overline{q}}=\int\frac{d^{3}p_{1}}{(2\pi)^{^{{\displaystyle 3}}}2E_{1}}\int\begin{array}{c}
{\displaystyle \frac{d^{3}p_{2}}{(2\pi)^{^{{\displaystyle 3}}}2E_{2}}}\end{array}\int\begin{array}{c}
{\displaystyle \frac{d^{3}p_{3}}{(2\pi)^{^{{\displaystyle 3}}}2E_{3}}}\end{array}\int\begin{array}{c}
\begin{array}{c}
{\displaystyle \frac{d^{3}p_{4}}{(2\pi)^{^{{\displaystyle 3}}}2E_{4}}(2\pi)^{4}}\end{array}\end{array}}\end{array}$}\\
\textcolor{black}{$\begin{array}{c}
{\displaystyle {\displaystyle \delta^{4}(p_{1}+p_{2}-p_{3}-p_{4})\Sigma|M_{s\overline{s}\rightarrow q\overline{q}}|^{{\displaystyle ^{2}}}(1-f_{q}(p_{1}))(1-f_{\overline{q}}(p_{2}))f_{s}(p_{3})f_{\bar{s}}}}(p_{4})\end{array}$}$\qquad$...(25b)\textcolor{black}{}\\
\textcolor{black}{Following {[} 8 ], we can say that there is only
one type of topologically distinct Feynmann diagram that contributes
towards the quark flavour changing process or the strange quark pair
production process. Evaluating it, performing trace calculations we
can find the squared matrix element. We basically follow the lines
of earlier works. Transforming variables as in the case before and
performing identical operations we can get the rate as }\\
\textcolor{black}{$\begin{array}{c}
{\displaystyle \begin{array}{c}
{\displaystyle R_{qg}=\frac{\alpha_{s}^{2}}{2\pi^{3}}T\int_{0}^{1}du\int_{0}^{1}dv\int_{0}^{1}dx\int_{0}^{1}dy\frac{u^{2}}{v}(1-\frac{4m_{i}^{2}}{s})^{\frac{1}{2}}(q_{0}^{2}-4m_{i}^{2})^{\frac{3}{2}}f_{Strange}f_{phase2}}\end{array}}\end{array}$}...(26)\textcolor{black}{}\\
\textcolor{black}{with}\\
\textcolor{black}{$\begin{array}{c}
{\displaystyle f_{Quarks}=f_{q}(\frac{q_{0}}{2}+p_{0})f_{\overline{q}}(\frac{q_{0}}{2}-p_{0})(1-f_{s}(\frac{q_{0}}{2}+p'_{0}))(1-f_{s}(\frac{q_{0}}{2}-p'_{0}))-}\end{array}$}\\
\textcolor{black}{$\begin{array}{c}
{\displaystyle \begin{array}{c}
{\displaystyle \qquad\qquad}\end{array}(1+f_{s}(\frac{q_{0}}{2}+p_{0}))(1+f_{s}(\frac{q_{0}}{2}-p_{0}))f_{q}(\frac{q_{0}}{2}+p'_{0})f_{\overline{q}}(\frac{q_{0}}{2}-p'_{0})}\end{array}$}\\
\textcolor{black}{$\begin{array}{c}
{\displaystyle f_{phase2}={\displaystyle [1+}[1-{\displaystyle \frac{4m_{i}^{2}}{s}}][{\displaystyle \frac{(1-x^{2})(1-y^{2})}{2}}+x^{2}y^{2}]]+\frac{4m_{i}^{2}}{s}}\end{array}$}\\
\textcolor{black}{}\\

\section{\textcolor{black}{\normalsize Results}\protect \\
\textcolor{black}{\normalsize }\protect \\
}

\textcolor{black}{The initial Conditions from the Self-Screened-Parton-Cascade
(SSPC) Model are as follows:}\\
\textcolor{black}{}\\
\textcolor{black}{}\begin{tabular}{|c|c|c|c|c|c|c|}
\hline 
 & $\begin{array}{c}
\tau_{{\displaystyle i}}(fm)\end{array}$ & $\begin{array}{c}
T(GeV)\end{array}$ & $\begin{array}{c}
{\displaystyle \lambda_{{\displaystyle g}}}\end{array}$ & $\begin{array}{c}
{\displaystyle {\displaystyle {\displaystyle \lambda_{{\displaystyle q}}}}}\end{array}$ &  & $\begin{array}{c}
{\displaystyle {\displaystyle \lambda_{{\displaystyle s}}}}\end{array}$\tabularnewline
\hline
\hline 
RHIC & 0.25 & 0.668 & 0.34 & 0.064 &  & 0.032\tabularnewline
\hline 
LHC & 0.25 & 1.02 & 0.43 & 0.086 &  & 0.043\tabularnewline
\hline
\end{tabular}\textcolor{black}{}\\
\textcolor{black}{In absence of known value of initial baryon number
density at the point when thermal equilibrium is attained (after which
point the system evolves according to the laws of hydrodynamics) we
perform a study with varying initial conditions. We vary the initial
baryon number density between 0.11 and 0.21/fm\textasciicircum{}3
while we change the initial quark to antiquark fugacity ratio between
1.1 and 1.9. Needless to say, most set of values would be of academic
interest only and not physically realisable. But, this could lead
to a better understanding, it is believed.}

We study comparatively the outputs with given initial baryon number
density and initial light quark to anti-quark fugacity ratio. We arrive
at the following conclusions :

1. The nature of variations of the physical quantities are more or
less in the line of earlier works. The \textcolor{black}{variations
of Temperature, Chemical Potential and Non equilibrium Fugacit}ies
for RHIC and LHC initial conditions are shown in figures 1 and 2 respectively.
\textcolor{black}{The decaying curves give temperature variations
while the positive rising curves are sequentially (from top) for gluon,
light quark, light anti-quark and strange quark fugacity variations.
Plots for both quark flavour changing process included and excluded
cases are shown. For identification please see point 4 below.As representative
values we take  baryon number density 0.15 and initial fugacity ratio
1.5.}

We observe that \textcolor{red}{\large }\\
\textcolor{black}{i) As expected, the temperature falls with time
while the non-equilibrium fugacities increase.}\\
\textcolor{black}{ii) Contrary to the Juttner case, the chemical potential
remains negative all along and as expected approaches zero as the
system equilibrates.}\\
\textcolor{black}{iii) The QGP,as expected remains to be gluon dominated.}\\
\textcolor{black}{iv) For a system at higher chemical potential, the
temperature falls at a slower rate signifying lesser amount of energy
expenditure to create partons which shows up in the slower rise of
all partons except for the light anti-quark, which shows a larger
growth rate due to the presence of the exponentiated chemical potential.}\textcolor{blue}{\large }\\
\\

2. For a given initial ratio, except for the light quarks and anti-quarks,
the output does not depend much on the initial baryon number density.
For the light quarks and antiquarks this variation is due to the presence
of the exponentiated non-zero chemical potentials. \\

3. For a given initial baryon number density we can recast the equation
for baryon number density in the form \[
e^{\mu_{q}/T}=\frac{C}{2\lambda_{q}}+\sqrt{\frac{1}{D}+\frac{C^{2}}{4\lambda_{q}^{2}}}\]
\\
$,\qquad\qquad\qquad\qquad\qquad\qquad\qquad\qquad\qquad$...(27)

where C is a constant for a given temperature, light quark fugacity
and given baryon number density. Here D is the light quark to antiquark
initial fugacity ratio. For a fixed light quark fugacity and temperature,
as D falls clearly the RHS of the above equation increases which indicates
a rise in the chemical potential. Again for a system of higher chemical
potential, the temperature has to drop at a slower rate due to the
constraint imposed. Hence we observe that\\
\\
i) The temperature falls at a slower rate for a smaller value of the
light quark to antiquark initial fugacity ratio.\\
ii) As the temperature falls at a slower rate, it would imply a lesser
expenditure of energy to produce particles in general . This would
show up in the slower rise of all fugacity values except the light
antiquark, which would show a higher growth rate. This is due to the
exponentiated chemical potential part. \\

4. For inclusion of the quark-flavour-changing process we observe
the following:\\
\\
 Due to the additional production of s-quarks from massless quarks
via qfcp, we see an additional increase in fugacity of the strange
quark while, the rate of equilibration falls for non-strange fermions
. 

5. \textcolor{black}{For some initial baryon number density and known
values of non-equilibrium fugacities and temperature switching distribution
functions would mean a shift in the value of the light quark chemical
potential, which is found by iteration. Since everything else remains
the same, to fit the baryon number density, clearly the chemical potential
is expected to 'pay the price'. This is exactly what happens. For
the MFD case, the chemical potential assumes a value much smaller
(-ve and also magnitudinally an order smaller) than that for the Juttner
case (in which case it is +ve). In the result for the number density
using the Juttner distributions we had an expression like $\begin{array}{c}
{\displaystyle \Sigma_{k=1}^{\infty}(-1)^{k-1}\{\frac{\lambda_{q}^{k}e^{\frac{k\mu_{q}}{T}}+\lambda_{\overline{q}}^{k}e^{-\frac{k\mu_{q}}{T}}}{k^{3}}\}}\end{array}$ . If we expand the exponential with an objective of suitable truncation,
keeping a 'k' factor in the numerator of the exponent with a much
larger value of the chemical potential, it would make the mathematics
still more complex as it would be required to consider many more number
of terms for the same accuracy. Therefore, we had to adopt the MFD
type of distribution functions ( with k=1 in the exponent ), where,
as required, we can easily use a truncated expansion of the exponential
term.} \\

\section{\textcolor{black}{\normalsize Acknowledgement}\protect \\
}

\textcolor{black}{The author gratefully acknowledges helpful discussions
with Prof. Bikash Sinha and Prof.Binayak Dutta Roy as also valuable
e-mail clarifications and encouragement from Profs. Z.J.He and Y.G.Ma
during the initial stages of the present work.}

\section{\textcolor{black}{\normalsize References}\protect \\
}

\textcolor{black}{1. N.Hammon et. al, Phys. Rev. C61, 014901 (1999)}\\
\textcolor{black}{2. STAR Collaboration, Phys. Lett. B567 (2003 )167
}\\
\textcolor{black}{3. BRAHMS Collaboration, Phys. Lett. B607 (2005)42}\\
\textcolor{black}{4. D.Dutta, A.K.Mohanty, K.Kumar, R.K.Choudhury,Phys.Rev.C60(1999)014905}\\
\textcolor{black}{5. D.Dutta, A.K.Mohanty, K.Kumar, R.K.Choudhury,Phys.Rev.C61(2000)064911}\\
\textcolor{black}{6. A.Sen, Rev. Bull. Cal. Math. Soc. 11 (1 \& 2
) 39-44 (2003)}\\
\textcolor{black}{7. Z.J.He, J.L.Long, Y.G.Ma, G.L.Ma B.Liu, Phys.
Rev. C69, 034901 (2004) }\\
\textcolor{black}{8. T.Matsui, B.Svetitsky, L.D.McLerran, Phys. Rev.
D34, 783 (1986)}\\
\textcolor{black}{9. Introduction to Relativistic Heavy Ion Collisions}\\
\textcolor{black}{Laszo P. Csernai, John Wiley and Sons, 1994}\\
\textcolor{black}{10.Li Xiong et.al., Phys.Rev.C49,2203 (1994)}\\
\textcolor{black}{11.F.A.Berends et.al, Phys. Lett. 103B,124(1981)}\\
\textcolor{black}{12. Fred Cooper et. al., www.arXiv.org/hep-ph/0207370}
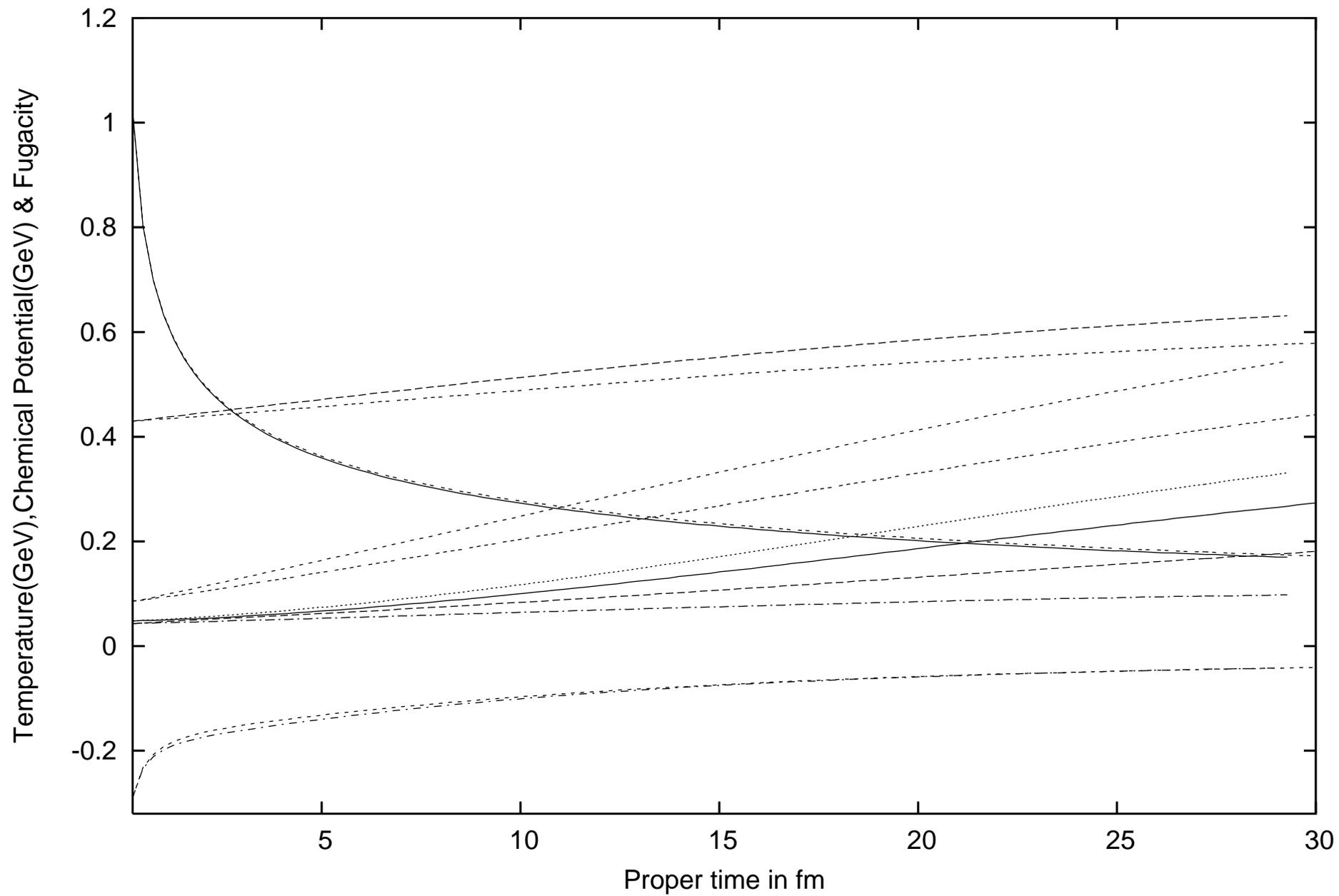
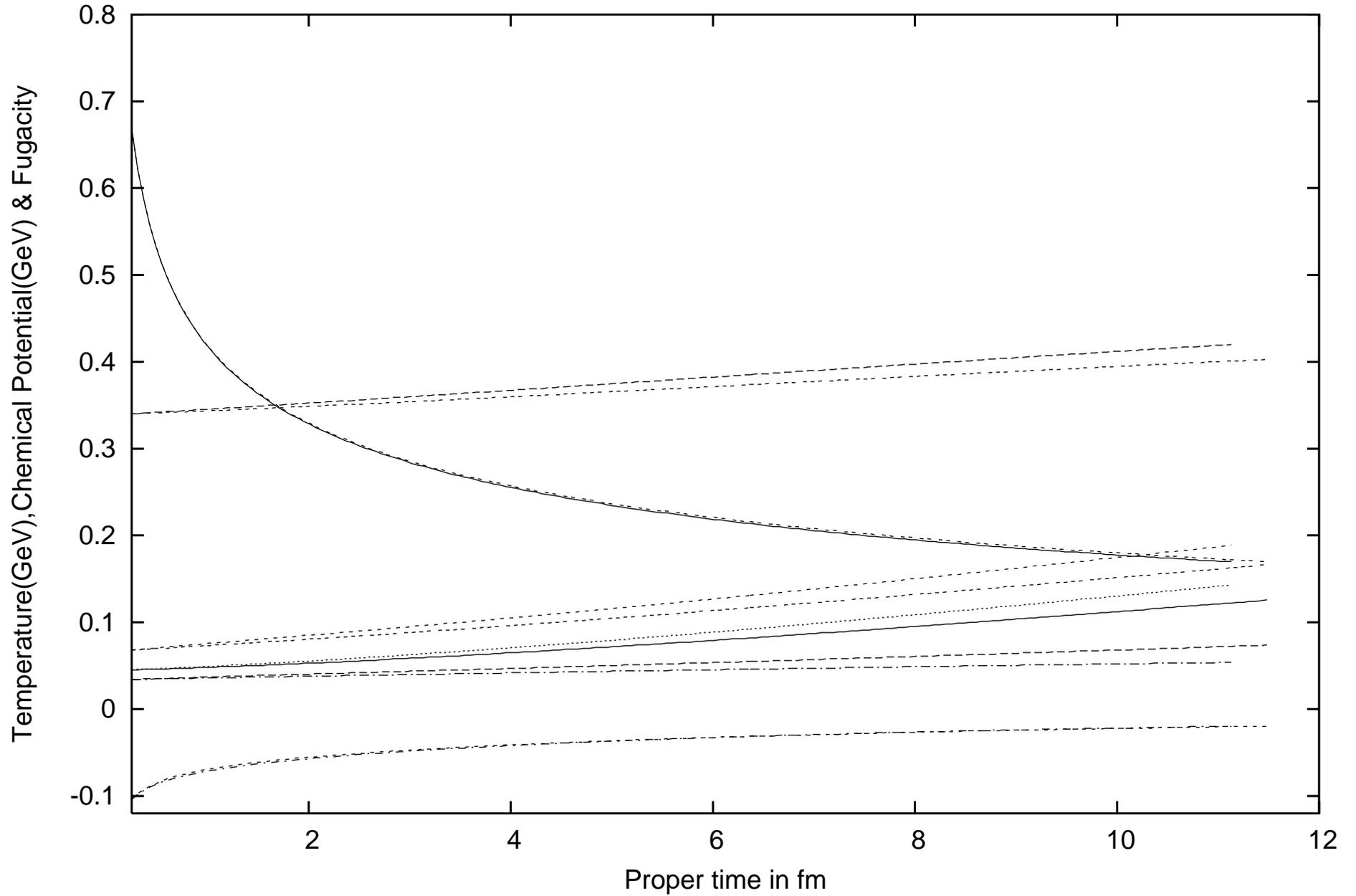
\end{document}